\documentclass[10pt,prl,aps,twocolumn,showpacs,superscriptaddress,floatfix]{revtex4}
\usepackage{graphicx}
\usepackage{amsmath}
\usepackage{epstopdf}
\usepackage{amsbsy}
\usepackage{color}

\begin{document}
\draft
\title{Universal Finite Temperature Properties of
a Three Dimensional Quantum Antiferromagnet
in the Vicinity of a  Quantum Critical Point}
\author{J. Oitmaa, Y. Kulik, and O.P. Sushkov }
\affiliation{School of Physics, The University of New South Wales,
  Sydney, NSW 2052, Australia}
\date{\today}
\begin{abstract}
We consider a 3-dimensional quantum antiferromagnet which can be driven
through a quantum critical point (QCP) by varying a tuning parameter g.
Starting from the magnetically ordered phase, the N{\'e}el temperature
will decrease to zero as the QCP is approached. From a generic quantum
field theory, together with numerical results from a specific
microscopic Heisenberg spin model, we demonstrate the existence of
universal behaviour near the QCP.  We compare our results 
with available data for $TlCuCl_3$.
\end{abstract}
\pacs{64.70.Tg
, 75.40.Gb
, 75.10.Jm
}

\maketitle
\newpage

The subject of continuous Quantum Phase Transitions (QPT's) and the behaviour of
quantum systems in the vicinity of the corresponding quantum critical
points is a frontier area of research both in theory and in experiment.
\cite{Sachdevphystoday,sachdev2010}
A QPT is a transition at zero temperature, in the nature of the ground
state, and is due to quantum fluctuations
that can be enhanced or suppressed by varying some coupling constant.
In real materials QPT's can be driven by pressure, by applied magnetic 
field, or by some other parameter.

In the present work we consider an O(3) QPT which occurs between a
magnetically ordered N{\'e}el phase and a magnetically disordered 
'valence-bond-solid' (VBS) phase in a class of SU(2) invariant Heisenberg spin
systems. This problem has attracted a great deal of attention in recent
years, mainly in two-dimensional (2D) systems. It has been established
that the interplay between quantum fluctuations and thermal fluctuations
at low but finite temperatures influences the dynamics in the vicinity
of a QPT in a highly nontrivial way \cite{Chak,Chub}. However, in 2D
systems there is no finite temperature magnetic order, due to the
well known Mermin-Wagner theorem. One would expect that in 3D systems
(3D + time)
the presence of a finite N{\'e}el temperature and an extended region of
magnetic order will affect the interplay between quantum and thermal
fluctuations, and lead to new features not seen in 2D. 
 An obvious
question is the nature of the vanishing of the N{\'e}el temperature and
its scaling with the magnetization and with the coupling constant
as the QPT is approached. To the best
of our knowledge the generic problem of the finite temperature behaviour
of 3D systems in the vicinity of an O(3) QPT has not been previously
considered. The present work addresses this question.

Specifically, we discuss three aspects of this question. The first is to
 consider a general Landau-Ginzburg field theory, 
which is independent of
the details of any microscopic model, and hence generic. The predictions
of this approach are then compared with experimental results for the
material TlCuCl${}_{3}$. Finally we present results obtained for a
specific microscopic Heisenberg spin model, obtained using a variety of
series-expansion methods. While the numerical precision close to the QPT
is only moderate, the results are consistent with the field theory
predictions, and reinforce our conclusion that the behaviour is universal.

To develop a quantum field theoretic description we start from the
standard 
 effective  Lagrangian describing an O(3) QPT, of the form~\cite{zinn,sachdev2010,Kulik}.
\begin{eqnarray} 
\label{LLs}
{\cal L}=\frac{1}{2}
\left({\dot{\vec \varphi}}-[{\vec \varphi}\times {\vec B}]\right)^2-
\frac{c^2}{2}\left({\bf \nabla}{\vec \varphi}\right)^2 
-\frac{m^2}{2}{\vec \varphi}^2
-\frac{\alpha}{4}[{\vec \varphi}^2]^2 
\end{eqnarray}
In the present work we consider zero magnetic field, $B=0$.
The vector field ${\vec \varphi}$ describes the staggered magnetization.
The QPT results from the mass term, assumed to be of the form 
$m^2=\lambda^2(g-g_c)$,
where $\lambda^2 > 0$ is a coefficient and
$g$ is a coupling parameter (In TlCuCl${}_{3}$ the coupling parameter
is an external hydrostatic pressure).
When $g > g_c$ the mass squared is positive and this corresponds to the magnetically 
disordered phase with gapped triply degenerate excitations. These are
sometimes called 'triplons' but we will use the term 'magnon' in both
phases. The zero temperature gap is
\begin{equation}
\label{gap}
\Delta=m=\lambda\sqrt{g-g_c} \ .
\end{equation}
When $g < g_c$ the mass squared is negative and  this results in a nonzero expectation value  
\begin{equation}
\label{vf}
|\langle{\vec \varphi}\rangle|=
\sqrt{\frac{|m^2|}{\alpha}}=\frac{\lambda}{\sqrt{\alpha}}\sqrt{g_c-g} \ .
\end{equation}
 that describes the spontaneous staggered magnetization at zero temperature.
This is a magnetically ordered phase with a gapped longitudinal mode and two 
transverse gapless Goldstone modes.
We note that $\varphi$ has dimensions of $energy^{-1/2}$, and therefore cannot
be directly compared with the dimensionless staggered magnetization.
The zero temperature energy of the magnetically ordered ground state is
\begin{equation}
\label{enf}
E=-\frac{\lambda^4}{4\alpha}(g-g_c)^2 \ .
\end{equation}
\begin{figure}[ht]
\vspace{10pt}
 \includegraphics[width=0.6\linewidth]{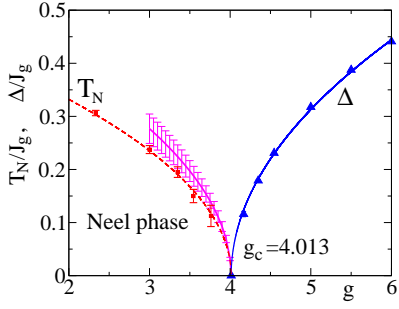}
 \caption{The phase diagram in the vicinity of a QPT.
The blue triangles connected by the blue solid line show results of 
series expansion calculations of the spin wave gap
in the magnetically disordered VBS phase.
The N{\'e}el  temperature  is shown at $g < g_c$ where the system is
magnetically ordered at $T < T_N$.
The red squares connected by the red dashed line show results of 
series expansion calculations of $T_N$.
The magenta solid line shows the field theory prediction for $T_N$.
}
 \label{tn}
\end{figure}
The generic phase diagram is shown in Fig.~\ref{tn}.
The specific parameter values shown in the figure correspond to the particular model
which we consider below.

\begin{figure}[ht]
 \includegraphics[width=0.2\linewidth]{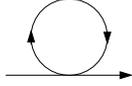}
 \caption{ Magnon selfenergy}
 \label{se}
\end{figure}

The magnetic ordering at $g < g_c$ is destroyed at $T > T_N$. To find $T_N$ we 
calculate the selfenergy $\Sigma$ shown in Fig.~\ref{se}. 
The four-leg vertex in Fig.~\ref{se}
is due to the quartic $\alpha$-term in (\ref{LLs}).
To calculate the single loop selfenergy in the magnetically disordered phase
it is sufficient to decouple the quartic interaction, 
$\alpha[{\vec \varphi}^2]^2 \to
\alpha \langle {\vec \varphi}^2\rangle {\vec \varphi}^2 \to
\Sigma {\vec \varphi}^2$.
When performing the decoupling, one has to be careful about the combinatorial
factor that is due to various ways of the field couplings.
A straightforward calculation gives the following selfenergy
in the magnetically disordered phase ($g > g_c$ or  $T>T_N$ at $g < g_c$),
\begin{equation}
\label{se1}
\Sigma=5\alpha\langle\varphi_i^2\rangle=5\alpha
\sum_{\bf k}\frac{1}{\omega_{\bf k}}
\left(n_{{\bf k}i}+\frac{1}{2}\right)\ ,
\end{equation}
where $i$ is any of three Cartesian components of ${\vec \varphi}$, and
$n_{{\bf k}i}=\langle a_{{\bf k}i}^{\dag}a_{{\bf k}i}\rangle
=1/(e^{\omega_{\bf k}/T}-1)$ is the thermal population of this component.
The quantum fluctuation part of (\ref{se1}) is ultraviolet divergent.
\begin{eqnarray}
5\alpha\sum_{\bf k}\frac{1}{2\omega_k}&=&\frac{5\alpha}{4\pi^2}
\int_0^{\Lambda}\frac{k^2dk}{\sqrt{\delta^2+c^2k^2}}\nonumber\\
&\approx& \frac{5\alpha}{8\pi^2 c^3}
\left[c^2\Lambda^2-\delta^2\ln\left(\frac{c\Lambda}{\delta}\right)\right] \ .
\nonumber
\end{eqnarray}
Here $\Lambda$ is an ultraviolet cutoff and $\delta$ is the gap in the 
spectrum, for example at $T=0$ and $g > g_c$, $\delta=\Delta$.
The quadratically divergent part $\propto \Lambda^2$ of the self energy
has to be removed by renormalization.  In other words, this part is absorbed in the 
value of the critical coupling constant $g_c$.
The logarithmic part depends on both the ultraviolet cutoff $\Lambda$
and the infrared cutoff $\delta$, and is therefore a real physical 
correction.
However, we expect this logarithmic correction to be small and therefore we disregard it,
(see also the discussion in Ref.~\cite{Kulik}).
The parameter that suppresses the correction is the prefactor $1/\pi^2$,
and in essence it is related to the 3D character of the problem.
Neither the existing experimental data presented below nor results of
numerical simulations also presented below have a sufficient accuracy to 
pin the logarithmic corrections down.
All in all this implies that the entire quantum part of the 
self energy is renormalized out, 
\begin{equation}
\label{se3}
\Sigma_R \approx 5\alpha\sum_{\bf k}\frac{n_{{\bf k}i}}{\omega_{\bf k}}\ ,
\end{equation}
where the subscript 'R' stands for 'renormalized'.
At $T=T_N$ the excitation spectrum is gapless, $\delta=0$,
$\omega_{\bf k}=ck$. Hence a calculation of the integral in (\ref{se3})
gives $\Sigma_R=\frac{5\alpha T_N^2}{12c^3}$.
If the magnon spectrum is anisotropic with three different principal
velocities then $c^3$ has to be replaced by $c_1c_2c_3$.
The magnon gap at the N{\'e}el temperature is zero, $\delta^2=m^2+\Sigma_R=0$, 
and hence
\begin{equation}
\label{TN}
T_N=\sqrt{\frac{12\lambda^2c_1c_2c_3}{5\alpha}}\sqrt{g_c-g}\ .
\end{equation}
Thus, the N{\'e}el temperature is directly proportional to the zero
temperature staggered magnetization (\ref{vf}). 
A similar scaling was obtained recently in Monte Carlo simulations
with various kinds of models~\cite{And}.

Equation (\ref{TN}) can be compared with experimental data~\cite{Ruegg}
for TlCuCl${}_{3}$.
\begin{figure}[ht]
\vspace{15pt}
 \includegraphics[width=0.6\linewidth]{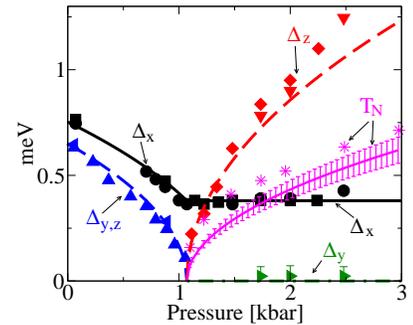}
 \caption{Zero temperature magnon  gaps and N{\'e}el temperature
in  TlCuCl${}_{3}$ versus pressure. All values are given in meV.
The system is magnetically disordered at $p\le p_c=1.07$kbar,
and magnetically ordered at $p\ge p_c$.
Symbols show experimental data from Ref.~\cite{Ruegg} and curves
show theoretical results.
}
\label{DT}
\end{figure}
Values of gaps at zero temperature versus pressure are plotted
in Fig.~\ref{DT}. The critical pressure is $p_c=1.07$kbar.
Note that Fig.~\ref{DT} is 'mirror reflected' compared to 
Fig.~\ref{tn},
the magnetically ordered phase is at $p > p_c$, therefore,
to compare with Eqs.(\ref{gap}),(\ref{vf}),(\ref{TN}) we choose
$g=-p$.
In the ideal situation corresponding to the action (\ref{LLs})
one should expect triply degenerate gapped excitations in the magnetically 
disordered phase at $p < p_c$,  as well as one longitudinal gapped mode
($\Delta _z$) and two gapless transverse modes in the magnetically
ordered phase at $p > p_c$.
In the real compound there is a small easy plane anisotropy and 
due to the anisotropy one
of the transverse magnons in the magnetically ordered phase is 
gapped, with $\Delta_x=0.38$meV. For the same reason the triple degeneracy 
at $p < p_c$ is lifted.
Disregarding the small anisotropy effects and using Eq.(\ref{gap})
we fit the gap in the magnetically disordered phase.
The fit is shown  in Fig.~\ref{DT} by the blue dashed line,
and results in the following value of  $\lambda$,
 $\lambda \approx 0.66\text{ meV/kbar}^{1/2}$.
Other parameters of the effective action (\ref{LLs})
were determined in the analysis of magnon spectra
and Bose condensation of magnons performed in Ref.~\cite{Kulik},
$c_1=7.09$meV, $c_2=1.12$meV, $c_3=0.51$meV, 
$\alpha=21(1\pm 0.2)\text{ meV}^{3}$.
Substitution to Eq.(\ref{TN}) gives the theoretical prediction of the 
N{\'e}el temperature plotted in Fig.~\ref{DT} by the solid magenta curve
with error bars that are mainly due to uncertainty in the value 
of $\alpha$.
This curve is very close to the experimental points, shown by magenta stars.
The experimental values of the N{\'e}el temperature are slightly higher
compared to the theory, especially close to the QPT point.
As one would expect the magnetic anisotropy, pointed out above,
leads to an enhancement of the  N{\'e}el temperature. We believe this
fully accounts for the discrepancy.

While the above theory is generic, and independent of the details of any
microscopic model, it is interesting and important to consider a specific model
and compare results with general theory. A specific microscopic model
can be analysed only numerically, so below we consider a sort of numerical
experiment versus the real experiment discussed above.
Many previous numerical studies of QPT's have been reported. These have
been largely based on Heisenberg spin models in which the system can be
tuned through a QPT by varying a particular coupling parameter in the
Hamiltonian. Most of these models have been two-dimensional. Examples
include antiferromagnets with strong and weak bonds, with or without
frustration \cite{singh1988,matsumoto2001,wenzel2009}, and bilayer
systems \cite{zheng1997,wang2006}, where the QPT separates a
conventional N{\'e}el  antiferromagnetic phase from a spin-dimerized
phase with only short range correlations and no magnetic order.

\begin{figure}[h]
 \includegraphics[width=0.4\linewidth]{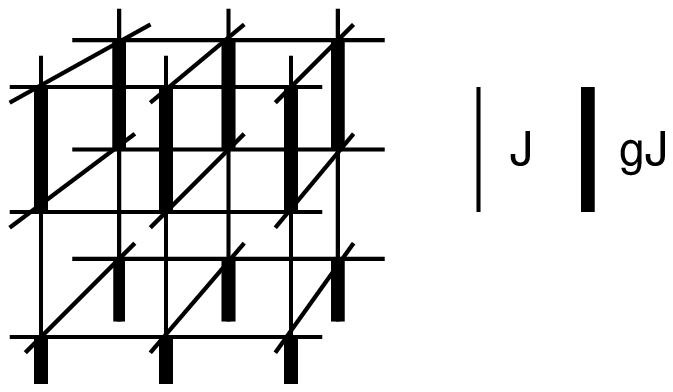}
\hspace{0.5cm}
 \includegraphics[width=0.5\linewidth]{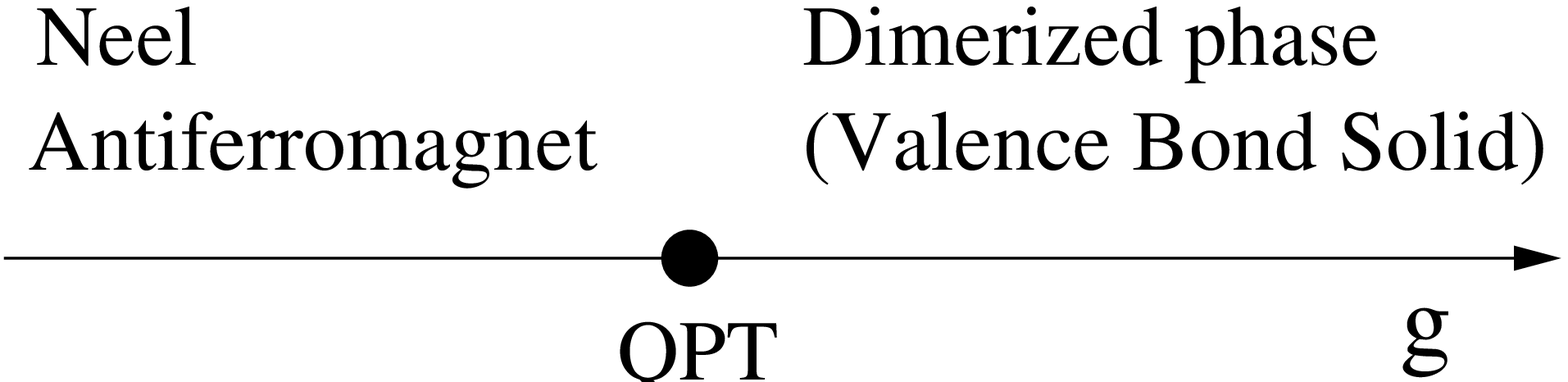}
 \caption{ (a) The model, with thin lines denoting $J$ bonds and thick
lines denoting $gJ$ bonds; (b) Schematic phase diagram of the model at
$T = 0$.
}
 \label{fig4}
\end{figure}

Here we consider a 3D spin-1/2 Heisenberg antiferromagnet. Our model,
shown in Fig.\ref{fig4}(a), has weak and strong bonds of strength $J$ and $gJ$
respectively. For $g=1$ we have an isotropic cubic antiferromagnet,
which has reduced staggered magnetization
in the ground state ($M_0$ = 0.42~\cite{oitmaa1994}) and a critical 
temperature $T/J$ = 1.89~\cite{oitmaa2004}. 
On the other hand, for $g >> 1$ the strong bonds
form spin-singlet dimers, leading to the VBS phase. A QPT
separates these phases, as shown schematically in Fig.\ref{fig4}(b). 
This model has been studied previously~\cite{nohadani2005} in connection
with magnetic-field induced QPT's, and the quantum critical point
was located at $g_c = 4.013 \pm 0.003$, using quantum Monte
Carlo (QMC) methods. However, the important questions of the universal
behaviour of the N{\'e}el temperature and the dynamics of the dimerized
phase were not discussed.

Our numerical calculations are based on series expansion methods
\cite{oitmaa2006}, and involve several separate parts. The various
series have been analysed in the usual way, via Pad\'e approximants. The
error bars shown on some of the data points are not statistical errors
but 'confidence limits' based on consistency and spread between different
high order approximants. For many data points, these error bars are
smaller than the point size.

We have used a 'dimer expansion' \cite{oitmaa2006} to obtain series for
the ground-state energy and the magnon energies in the VBS phase, in
powers of $1/g$, to orders 11 and 8 respectively. The latter provides a
direct series for the minimum gap at ${\bf k}=(\pi,\pi,0)$. Analysis of
the gap series has to allow for the expected square-root singularity at
$g_c$, and we have used a Huse transformation to remove this
singularity. The resulting gap data are shown in Fig.~\ref{tn}
by blue triangles. Our
estimate of the critical point $g_c$ obtained from this data is fully
consistent with, although somewhat less precise than the Monte Carlo estimate 
$g_c = 4.013$. We use this value in our further analysis. The gap data
can be very well fitted by the expression $0.316\sqrt{g-g_c}$, which is
shown in Fig.~\ref{tn} by the blue solid line. This provides the
estimate $\lambda= 0.316J_g$, where
\begin{displaymath}
 J_g = J(1+g)
\end{displaymath}
is an average exchange parameter, used hereafter to set an energy scale.

Results for the magnon energies near ${\bf k}=(\pi,\pi,0)$, fitted to
the expression
\begin{displaymath}
      \epsilon({\bf k}) = \sqrt{\Delta ^2 + c_1^2(\pi-k_1)^2 +
c_2^2(\pi-k_2)^2 + c_3^2k_3^2}
\end{displaymath}
provide estimates of the magnon velocities near the QCP,
$c_1=c_2=0.516J_g, c_3=0.337J_g$. These values contain an
uncertainty up to $\pm 5\%$.
 
Next, we have used 'Ising expansions' \cite{oitmaa2006} in the N{\'e}el
phase to obtain series for the ground state energy and magnetization to
order 12 in an anisotropy parameter $x$. These energy series,
evaluated at $x$ via Pad{\'e} approximants, provide the data
shown in Fig.5 by red squares. The energies in the VBS phase, discussed
above, are shown as black circles. As can be seen, the two energy
curves, from the N{\'e}el and VBS phases respectively, meet smoothly at
the QCP, as expected for a second-order transition. 
\begin{figure}[h]
\vspace{10pt}
 \includegraphics[width=0.6\linewidth]{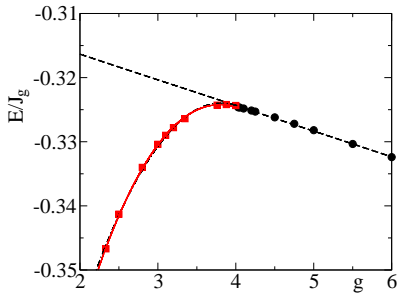}
 \caption{ The ground-state energy in the N{\'e}el phase (red squares)
and dimer phase (black circles) versus the coupling constant. The
lines are fits to the energy, as discussed below.}
 \label{en}
\end{figure}
This data can then
be used to estimate the parameter $\alpha$ in Eq.(\ref{enf}).
The VBS energy can be
accurately fitted with a straight line $E_{VBS}/J_g = -0.3244 -
0.0040(g-g_c)$, shown as the black dashed line in Fig.~\ref{en}. The N{\'e}el
data can be fitted with a quadratic expression, as in Eq.(\ref{enf}),
$E_N/J_g=E_{VBS}(g)+0.00003-0.0101(g-g_c)^2$.
However this fitting is subject to uncertainty, as the energies near the QCP 
are only changing in the 4th figure, and the data are not that precise. Indeed,
inclusion of a small cubic term in the fit changes the coefficient of
the quadratic term significantly,
$E_N/J_g=E_{VBS}(g)-0.00028-0.0088(g-g_c)^2+0.0008(g-g_c)^3$.
Comparing the coefficient of the quadratic term with Eq.(\ref{enf})
we determine the value of the quartic coupling constant.
 Our final estimate is $\alpha = 0.283
J_g^3$, with an uncertainty of $\pm 15\%$.

Substitution of the determined parameters into Eq.(\ref{TN}) gives
$T_N/J_g = 0.275\sqrt{g_c-g}$. This dependence is shown in Fig.~\ref{tn}
by the solid magenta line with error bars.
The main uncertainty, $\sim 7\%$, in the  coefficient 0.275
 comes from the uncertainly in the value of 
$\alpha$ discussed above. An additional few per cent come from
uncertainties in $\lambda$ and magnon velocities.
Altogether we estimate the computational uncertainly in the value
of the coefficient 0.275 as 10\%. This is shown as error bars in the
solid magenta curve in Fig.~\ref{tn}.\\

Finally, we compute high-temperature expansions for the N{\'e}el
susceptibility. This is the response to a 'staggered' field. 
This susceptibility is expected to 
have a strong divergence at the critical temperature and can be used to
estimate $T_{N}(g)$. The Ne\'el temperature calculated in this way is shown in
Fig.~\ref{tn} by the red squares. The red dashed line just connects the data
points for guidance.
The agreement between prediction of the universal theory shown by the magenta
curve and results of the series computations is quite satisfactory.

In summary, we have shown that a 3-dimensional antiferromagnet in the
vicinity of a quantum critical point is expected to show universal
behaviour, including scaling of the N{\'e}el temperature with the ground
state magnetization and with the coupling constant. We predict the universal 
scaling. Our prediction based on a field theory
 accurately describes recent data on the material $TlCuCl_3$. 
The universal prediction is supported by numerical results obtained
for a microscopic S=1/2 Heisenberg spin model with strong and weak
bonds, which is a specific example of a 3D antiferromagnet with a QCP.
Results are obtained via a variety of series-expansion calculations, and
are shown to be in reasonable agreement with the predicted universal
behaviour, within numerical uncertainties.

We thank Anders Sandvik and Songbo Jin for helpful discussions and for
providing us with their data before publication.
We also thank  Cristian Batista for important comments.
Computing resources were provided by the Australian 
Partnership for Advanced Computing (APAC) National Facility.

\end{document}